\documentclass[aps,twocolumn,superscriptaddress,prb,longbibliography]{revtex4-2}

\usepackage[utf8]{inputenc}
\usepackage[T1]{fontenc}
\usepackage{dcolumn}
\usepackage{bm}
\usepackage{amsmath}
\usepackage{natbib}
\usepackage{graphicx} 
\usepackage{tabularx}
\usepackage{multirow}
\usepackage{rotating} 
\usepackage{makecell}
\usepackage{parskip}
\usepackage{amssymb}
\usepackage{textgreek}
\usepackage[normalem]{ulem}
\usepackage{hyperref}
\hypersetup{colorlinks=true, linkcolor=blue, citecolor=blue, urlcolor=blue}

\usepackage{float}

\usepackage{lineno}

\newcommand{\bea}{\begin{eqnarray}}
\newcommand{\eea}{\end{eqnarray}}
\DeclareUnicodeCharacter{2062}{}

\begin{document}

\title{Probing the superconducting gap structure of ScRuSi via $\mu$SR and first-principles calculations}
\author{K. Panda}
\affiliation{Department of Physics, Ariel University, Ariel 40700, Israel} 
\affiliation{Department of Physics, Ramakrishna Mission Vivekananda Educational and Research Institute, Belur Math, Howrah 711202, West Bengal, India} 
\author{A. Bhattacharyya}
\email{Corresponding author: amitava.bhattacharyya@rkmvu.ac.in} 
\affiliation{Department of Physics, Ramakrishna Mission Vivekananda Educational and Research Institute, Belur Math, Howrah 711202, West Bengal, India}
\author{P. N. Ferreira}
\email{Corresponding author: pedroferreira@usp.br} 
\affiliation{Universidade de S\~ao Paulo, Escola de Engenharia de Lorena, DEMAR, 12612-550, Lorena, Brazil}
\affiliation{Institute of Theoretical and Computational Physics, Graz University of Technology, NAWI Graz, 8010 Graz, Austria}
\author{Rajib Mondal}
\affiliation{UGC-DAE Consortium for Scientific Research, Kolkata Centre, Bidhannagar, Kolkata 700 106, India}
\author{A Thamizhavel}
\affiliation{Department of Condensed Matter Physics and Materials Science, Tata Institute of Fundamental Research, Homi Bhabha Road, Colaba, Mumbai 400 005, India}
\author{D. T. Adroja} 
\affiliation{ISIS Neutron and Muon Source, Rutherford Appleton Laboratory, Chilton, Didcot Oxon, OX11 0QX, United Kingdom} 
\affiliation{Highly Correlated Matter Research Group, Physics Department, University of Johannesburg, PO Box 524, Auckland Park 2006, South Africa}
\author{C. Heil}
\affiliation{Institute of Theoretical and Computational Physics, Graz University of Technology, NAWI Graz, 8010 Graz, Austria}
\author{L. T. F. Eleno}
\affiliation{Universidade de S\~ao Paulo, Escola de Engenharia de Lorena, DEMAR, 12612-550, Lorena, Brazil}
\author{A. D. Hillier}
\affiliation{ISIS Neutron and Muon Source, Rutherford Appleton Laboratory, Chilton, Didcot Oxon, OX11 0QX, United Kingdom}

\date{\today}
\begin{abstract}
 
In this study, we present a thorough investigation into the superconducting state of the ruthenium-based ternary equiatomic compound ScRuSi. Our analysis combines experimental techniques, including muon spin rotation/relaxation ($\mu$SR) and low-temperature resistivity measurements, with theoretical insights derived from first-principles calculations. The low-temperature resistivity measurements reveal a distinct superconducting phase transition in the orthorhombic structure of ScRuSi at a critical temperature ($T_\text{C}$) of $2.5$\,K. Further, the TF-$\mu$SR analysis yields a gap-to-critical-temperature ratio of $2\Delta/k_\mathrm{B}T_\mathrm{C} = 2.71$, a value consistent with results obtained from previous heat capacity measurements. The temperature dependence of the superconducting normalized depolarization rate is fully described by the isotropic $s$-wave gap model. Additionally, zero-field $\mu$SR measurements indicate that the relaxation rate remains nearly identical below and above $T_\text{C}$. This observation strongly suggests the preservation of time-reversal symmetry within the superconducting state. By employing the McMillan-Allen-Dynes equation, we calculate a $T_\text{C}$ of $2.11$\,K from first-principles calculations within the density functional theory framework. This calculated value aligns closely with the experimentally determined critical temperature. {The coupling between the low-frequency phonon modes and the transition metal d-orbital states play an important role in governing the superconducting pairing in ScRuSi.} The combination of experimental and theoretical approaches provides a comprehensive microscopic understanding of the superconducting nature of ScRuSi, offering insights into its critical temperature, pairing symmetry, and the underlying electron-phonon coupling mechanism.

\end{abstract}
\pacs{71.20.Be, 75.10.Lp, 76.75.+i}

\maketitle

\section{Introduction}

Ternary equiatomic intermetallic superconductors, characterized by the general formula RTX, where R is a transition or rare earth metal, T is a $4d$ or $3d$ transition metal, and X belongs to group IV or V, have become a focal point of research. {This heightened interest arises from the potential of these materials to serve as a promising platform for exploring unconventional superconductivity. One topical example is the centrosymmetric (Ta,Nb)OsSi, which has been reported to exhibit time-reversal symmetry (TRS) breaking while concurrently displaying a fully gapped conventional superconductivity~\cite{Ghosh2022}.

In recent years, there has been a notable increase in the exploration of Ru-based ternary equiatomic intermetallic superconductors~\cite{Shang2022,Das2021,Barz1980,Shirotani1995,Meisner1983,Shirotani1993,WongNg2003}. A particular focus has been placed on the noncentrosymmetric (Ta,Nb)RuSi compounds, {which are characterized by a substantial antisymmetric spin-orbit coupling and hourglass-like electronic dispersions, reminiscent of those observed in three-dimensional Kramers nodal-line semimetals~\cite{Shang2022}. Muon-spin rotation and relaxation ($\mu$SR) measurements have provided valuable insights into the superconducting properties of (Ta,Nb)RuSi, suggesting that these compounds are fully gapped time-reversal symmetry-breaking superconductors~\cite{Shang2022}. Beyond (Ta,Nb)RuSi, other Ru-based ternary equiatomic superconductors have also drawn a significant attention. Examples include $h$-(Zr,Hf,Ti)RuP~\cite{Barz1980}, $h$-ZrRuSi~\cite{Shirotani1995}, $h$-ZrRuAs ($T_\mathrm{C}$ $\approx$ 12\,K)~\cite{Meisner1983}, and $h$-ZrRuP~\cite{Shirotani1993}. Notably, MoRuP stands out for having the highest critical temperature ($T_\mathrm{C}$) among the ternary equiatomic intermetallic RTX superconductors, reaching 15.5\,K~\cite{WongNg2003}.}

\begin{figure*}[t]
\centering
\includegraphics[width=0.9\linewidth]{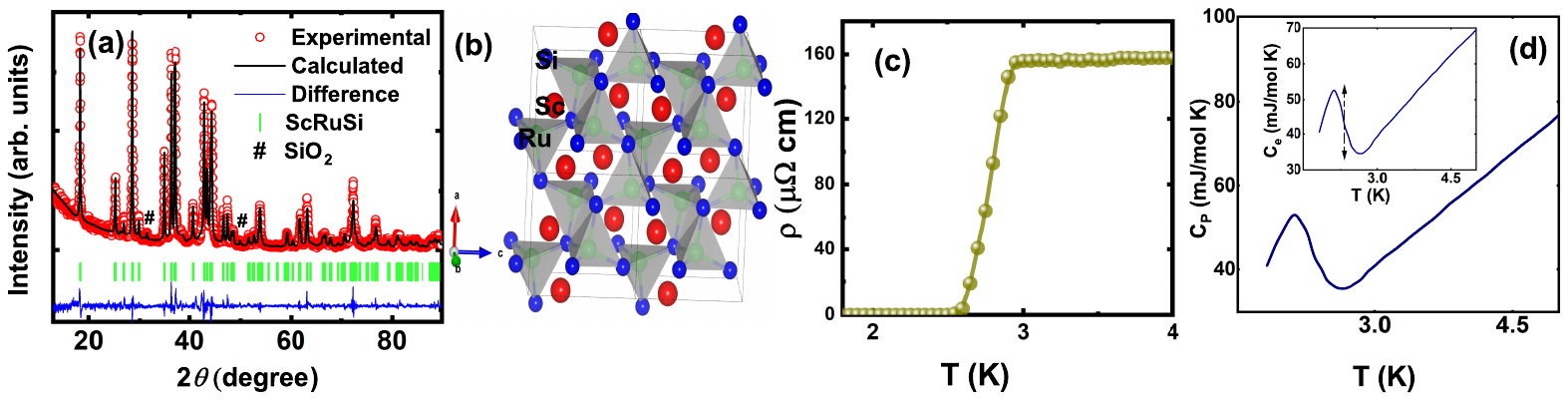}\hfil
\caption{(a) The powder x-ray diffraction pattern of ScRuSi subjected to Rietveld refinement is illustrated, with the experimental data represented by red circles and the refined result depicted as a solid black line. The tetragonal phase of ScRuSi, characterized by space group $Pnma$ (No.~62), serves as the primary phase in the refinement. Bragg peak positions are denoted by vertical green bars, and the difference between theoretical and experimental data is shown by the blue line. (b) The crystal structure of ScRuSi in the orthorhombic phase is displayed, featuring Sc (in large size, red), Ru (medium size, green), and Si atoms (small size, blue). (c) The resistivity of ScRuSi exhibits a low-temperature dependence, reaching a zero resistivity state at $T_\mathrm{C}$ = 2.5\,K. \textcolor{black}{(d) Temperature dependence of heat capacity, $C_{P}$(T), for measured in zero applied field. Inset shows the electronic contribution to the zero-field heat capacity, $C_{e}$, as a function of temperature.}}
\label{fig1}
\end{figure*}
 
These intermetallic compounds exhibit a diverse range of structural prototypes, typically crystallizing in four distinct forms: $(i)$ PbClF-type ($P4/nmm$, no.~129)~\cite{Welter1993}, $(ii)$ orthorhombic TiNiSi-type ($Pnma$, no.~62)~\cite{Morozkin1999}, and the noncentrosymmetric $(iii)$ hexagonal ZrNiAl-type ($P\bar{6}2m$, no.~189)~\cite{Barz1980}, and $(iv)$ orthorhombic TiFeSi-type ($Ima2$, no.~46)~\cite{Shirotani1998}. Superconductivity has been reported in both the orthorhombic and hexagonal structures. {Typically, hexagonal phases exhibit higher critical temperatures ($T_{\text{C}}$) compared to orthorhombic phases, \emph{e.g.} isoelectronic $h$-ZrRuP ($T_{\text{C}}$ = 13\,K)~\cite{Shirotani1993} and $h$-MoNiP ($T_{\text{C}}$ = 13\,K)~\cite{Shirotani2001}. However, there are exceptions to this trend, as $o$-MoRuP~\cite{WongNg2003}, with a remarkably high $T_{\text{C}}$ of 15.5\,K. }

The emergence of superconductivity in ScRuSi is particularly noteworthy given that neither Sc nor Si are known superconductors at ambient pressure, and pure bulk Ru displays superconductivity only at 0.47\,K in its hexagonal form~\cite{goodman1951two,Hulm1957}. Moreover, Ru is recognized as the fourth ferromagnetic chemical species at ambient temperature~\cite{Quarterman2018}. Ruan~{\it et al.} reported the occurrence of zero resistivity below $T_\mathrm{C} = 2.5$~K in the orthorhombic phase of ScRuSi~\cite{Ruan2016}. It is intriguing to note that no superconductivity was detected in the hexagonal phase down to 1.8\,K.  \emph{Ab initio} calculations have highlighted the involvement of low-frequency phonon modes, attributed to the transition-metal atoms. This contrasts with the high-frequency modes arising from the vibrations of Si atoms, suggesting a nuanced interplay of electronic and vibrational contributions to the superconducting behavior of ScRuSi~\cite{Uzunok2020}.

In recent studies, we have reported the presence of fully gapped nodeless BCS superconductivity in ZrIrSi and HfIrSi~\cite{Panda2019,Bhattacharyya2019}. Given that Sc shares structural similarities with Hf and Zr due to its chemical environment, investigating the nature of superconductivity in $o$-ScRuSi becomes imperative. In this regard, a systematic $\mu$SR study has been conducted to probe the microscopic aspects of the superconducting order parameter in $o$-ScRuSi. The analysis of the superfluid density at low temperatures reveals a saturating behavior, indicative of the fully gapped superconducting nature of ScRuSi, with a calculated $2\Delta/k_{B}T_{C}$ ratio of 2.71. This observation aligns with the characteristics of a conventional BCS superconductor. Additionally, the equality of the relaxation rates in the zero-field $\mu$SR asymmetry spectra above and below $T_\mathrm{C}$ further confirms the preservation of time-reversal symmetry within the superconducting state. To provide a theoretical framework for the superconducting state, first-principles calculations within the semi-empirical McMillan-Allen-Dynes theory~\cite{allen1975} have been employed. The results demonstrate that superconductivity in ScRuSi, in contrast to some of its counterparts, can be explained according to the conventional superconducting pairing mechanism.

\begin{table*}
\caption{Atomic positions, occupancy, and atomic displacement parameters obtained through the Rietveld refinement of the powder XRD data for ScRuSi. The crystal structure belongs to space group $Pnma$, with a unit cell volume of $190.8002\,\AA^{3}$. The refinement results include a $\chi^{2}$ value of 2.26\,\%, a Bragg R-factor of 4.10\,\%, and an R$_{f}$ value of 3.18\,\%.}
\begin{tabular}{m{1.5cm} m{3cm} m{2.5cm} m{2.5cm} m{2.5cm} m{2.5cm} m{2.5cm}}
      \hline 
      \hline
     Atom & Wyckoff symbol & x & y & z & Occupancy & B$_{iso}$ (\AA$^{2}$)\\
     \hline
     Sc & 4c & -0.0035(3) & 0.25 & 0.6869(4) & 1 (fixed) & 0.0514(3)\\
     Ru & 4c & 0.1588(7) & 0.25 & 0.0597(8) & 1 (fixed) & 0.0533(8)\\
     Si & 4c & 0.2919(8) & 0.25 & 0.3864(5) & 1 (fixed) & 0.1266(6)\\
      \hline
\end{tabular}
\label{tab:xrd}
\end{table*}

\section{Methods}

\subsection{Experiment}

In this investigation, a polycrystalline sample of ScRuSi was synthesized through a standard arc melting process conducted on a water-cooled copper hearth. The synthesis involved using high-purity Sc, Ru, and Si elements, each with a purity exceeding 99.9\%, in a stoichiometric ratio. To ensure phase homogeneity, the arc-melted ingot underwent multiple remelting cycles. The structural characterization of the specimen was performed via X-ray diffraction using a Panalytical X-ray diffractometer equipped with Cu-$K_{\mathrm{\alpha}}$ radiation ($\lambda$ = 1.540596 \AA) at ambient temperature. DC-electrical resistivity measurements were conducted using a physical property measurement system (PPMS, Quantum Design). The standard dc-four probe technique was employed for these measurements, covering a temperature range down to 1.8\,K. \textcolor{black}{The heat capacity was measured using the Quantum Design Physical Properties Measurement System (PPMS).} 

We have performed $\mu$SR measurements using the MuSR spectrometer at the ISIS Pulsed Neutron and Muon Source in UK. The measurements, conducted at low temperatures down to 0.1 K, employed a dilution refrigerator in both longitudinal and transverse directions. The powdered sample was mounted on a high-purity silver plate (99.995\%) using GE varnish and wrapped with silver foil. The technique involved implanting 100\% spin-polarized positive muons ($\mu^{+}$) into the sample for the measurements.

In the $\mu$SR measurements for ScRuSi, the muon spin undergoes rotation and relaxation in response to local magnetic fields. Each $\mu^{+}$ decays with a mean lifetime of $2.2~\mu$s, emitting a positron. The emitted positrons are preferentially directed based on the orientation of the muon spin vector and are detected by both forward (F) and backward (B) detectors. For zero-field (ZF) measurements, the detectors are aligned in the longitudinal direction (parallel to the muon spin), and the asymmetry ($A(t)$) between the counts of positrons detected in the forward detectors ($N_{\mathrm{F}}(t)$) and backward detectors ($N_{\mathrm{B}}(t)$) is given by the expression:
\[ A(t) = \frac{N_{\mathrm{F}}(t) - \alpha N_{\mathrm{B}}(t)}{N_{\mathrm{F}}(t) + \alpha N_{\mathrm{B}}(t)}, \]
where $\alpha$ represents an instrumental calibration constant determined in the normal state under a small transverse magnetic field of approximately 2\,mT. To achieve zero field, any stray magnetic fields originating from the Earth or neighboring instruments were minimized to within a 1 $\mu$T limit. This was accomplished using an active compensation system with three pairs of correction coils. All transverse field (TF)-$\mu$SR experiments were performed after field-cooling the sample with an applied magnetic field of 30\,mT. {The TF-$\mu$SR data were collected at various temperatures ranging from 0.1 to 3.5\,K}. The $\mu$SR asymmetry spectra  were analyzed using the Windows Muon Data Analysis (WiMDA) open-source software~\cite{Pratt2000}.

\subsection{Theory}

Electronic and vibrational properties were computed using the \textsc{Quantum Espresso}~\cite{QE2, QE3} suite employing scalar-relativistic optimized norm-conserving Vanderbilt pseudopotentials (ONCV)~\cite{ONCV1, ONCV2} and the Perdew-Burke-Ernzerhof parametrization within the generalized gradient approximation for the exchange and correlation functional~\cite{PBE}. Kohn-Sham orbitals were expanded on plane waves with a kinetic energy cutoff of 50\,Ry for the wavefunctions and 200\,Ry for the charge density. An unshifted 8$\times$8$\times$8 $\mathbf{k}$-grid sampling over the Brillouin zone was adopted according to the Monkhorst-Pack scheme \cite{k-mesh}. Self-consistent-field (SCF) calculations considering a $10^{-10}$\,Ry convergence threshold was carried out using Methfessel-Paxton gaussian smearing \cite{MP-smearing} with a spreading of 0.01\,Ry for Brillouin-zone integration. The density of states and Fermi surfaces were obtained over 40$\times$40$\times$40 k-points. The experimental lattice parameters and atomic positions were adopted. 

Phonon frequencies were obtained by Fourier interpolation of the dynamical matrices on 4$\times$4$\times$4 $\mathbf{q}$-grid within Density Functional Perturbation Theory (DFPT) \cite{DFPT} adopting a threshold for self-consistency of 10$^{-14}$\,Ry. Electron-phonon properties were computed on 32$\times$32$\times$32 $\mathbf{k}$-grid self-consistently within the linear response theory \cite{DFPT}.

\section{Results and Discussion}

\subsection{Structural characterization and Physical properties}
 
To check how pure the ScRuSi sample is, we looked at its x-ray diffraction patterns at room temperature. We used the Rietveld refinement method with \textsc{FULLPROF} software~\cite{RodrguezCarvajal1993} for the analysis. The refined x-ray pattern, as illustrated in Figure~\ref{fig1}a, exhibits a remarkable agreement between the theoretical and experimental data, as reflected in a low $\chi^{2}$ value of 2.26\,\%. The agreement confirms that the obtained compound follows the crystalline structure of the orthogonal space group $Pnma$. The refined structural parameters are $a = 6.6154$\,\AA, $b$ = 4.0932\,\AA, and $c$ = 7.0462\,\AA, along with a unit cell volume of 190.8\,\AA$^{3}$. These values align closely with the parameters reported in Ref.~\cite{Ruan2016}. Table~\ref{tab:xrd} provides a comprehensive overview of the atomic positions, occupancy, and atomic displacement parameters obtained from the Rietveld refinement of the powder X-ray diffraction (XRD) data for ScRuSi. {In Figure~\ref{fig1}a, we observe impurity peaks from SiO$_{2}$, denoted with \#. Our estimation suggests approximately 3\% impurity exists in the current sample.  This impurity, being nonmagnetic, does not exert any influence on the findings concerning the superconducting properties of ScRuSi as presented in this study. A schematic of the unit cell obtained from the Rietveld analysis of the XRD data of ScRuSi  is shown in the Fig. \ref{fig1}(b).}

Superconductivity is confirmed by looking at the resistivity behavior. The temperature dependence of the electrical resistivity, $\rho(T)$, in the absence of an applied magnetic field at ambient pressure, is depicted in Figure \ref{fig1}c. Notably, the electrical resistivity data exhibits a distinct superconducting phase transition at $T_{\text{C}}$ = 2.5\,K. In the normal state, before the onset of superconductivity, the resistivity data demonstrates a characteristic metallic behavior. \textcolor{black}{Heat capacity, C$_\text{P}$, as a function of temperature is shown in figure \ref{fig1}(d) in zero applied field. Above $T_{\text{C}}$ in the normal state, the C$_\text{P}$(T) data can be described using $C_\text{P}(T) = \gamma T +\beta T^{3}$, where $\gamma T$ is the electronic heat capacity coefficient, and $\beta T^{3}$ is the lattice (phonon) contribution to the specific heat. Fitting gives $\gamma$ = 12.92 mJ mol$^{-1}$ K$^{-2}$ and $\beta$ = 0.095 mJ mol$^{-1}$K$^{-4}$. Inset of the Figure \ref{fig1}(d) shows the temperature dependence of the electronic specific heat, $C_\text{e}$ (T), obtained by subtracting the phonon contribution from $C_\text{P}(T)$. The jump in the heat capacity at $T_{\text{C}}$ is estimated to be $\Delta C_{e}/\gamma T_{C}$ = 1.31, which is close to 1.43 expected for weak-coupling BCS superconductors~\cite{Bardeen1957}.}
 
\begin{figure}[tb]
\centering
\includegraphics[width=0.9\linewidth]{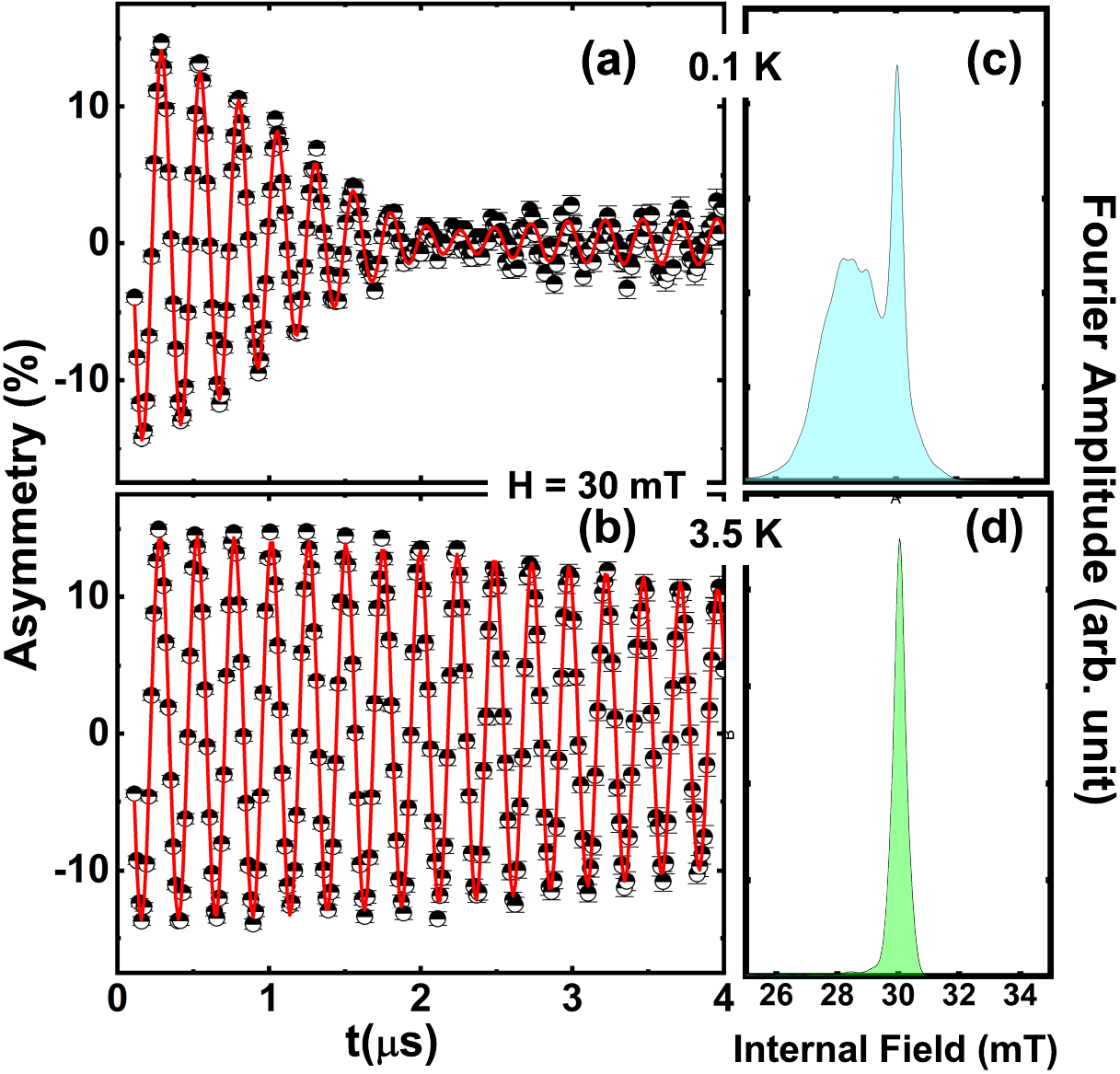}\hfil
\caption{Transverse-field $\mu$SR asymmetry spectra  for ScRuSi obtained in the field-cooled state under an applied magnetic field of 30\,mT at two different temperatures: (a) 0.1\,K and (b) 3.5\,K. Alongside these asymmetry spectra  are their respective Fourier-transformed maximum entropy spectra in (c) and (d). The solid red curves in figure (a) and (b) depict the fits to the data points using the functional form described by Eq. \ref{TF}, as elaborated in the text. The pronounced fast damping of the signal observed in the superconducting state (0.1\,K) reflects the inhomogeneous field distribution experienced by the muon ensemble due to the formation of vortices. This phenomenon provides crucial insights into the superconducting behavior of ScRuSi under external magnetic fields.}
\label{fig2}
\end{figure}

\begin{figure}[t]
\centering
\includegraphics[width=0.9\linewidth]{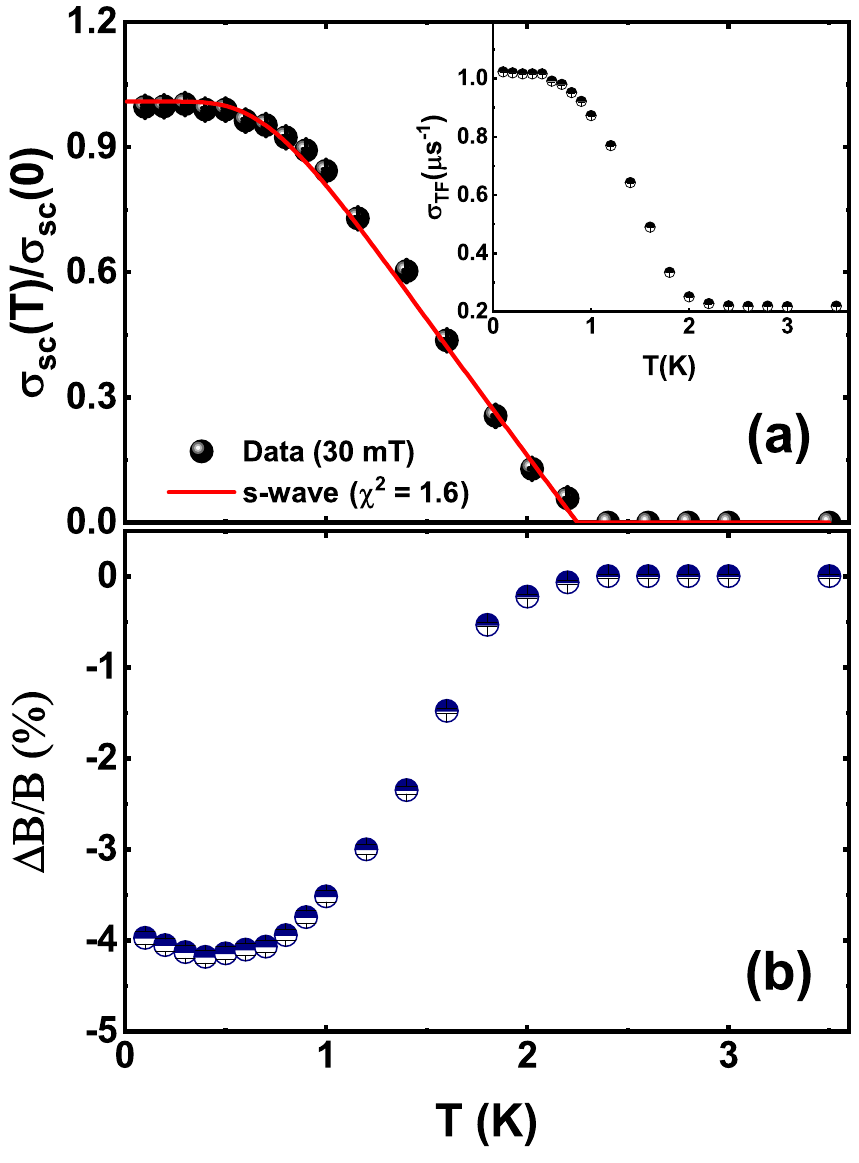}\hfil
\caption{Key results related to the superconducting behavior of ScRuSi under an applied magnetic field of 30\,mT. In (a), the temperature dependence of the superconducting normalized depolarization rate, $\sigma_\mathrm{sc}(T)/\sigma_\mathrm{sc}(0)$, is presented. The solid red line represents a fit using the $s$-wave gap structure model. The inset displays the temperature dependence of the total muon-spin depolarization rate, $\sigma_\mathrm{T}$. Moving to (b), the figure shows the temperature dependence of $\Delta B = (B_\mathrm{sc} - B_\mathrm{app})/B_\mathrm{app}$ field shifts, expressed as a percentage of the applied field $B_\mathrm{app}$. }

\label{fig3}
\end{figure}

\begin{figure}[t]
\centering
\includegraphics[width=0.9\linewidth]{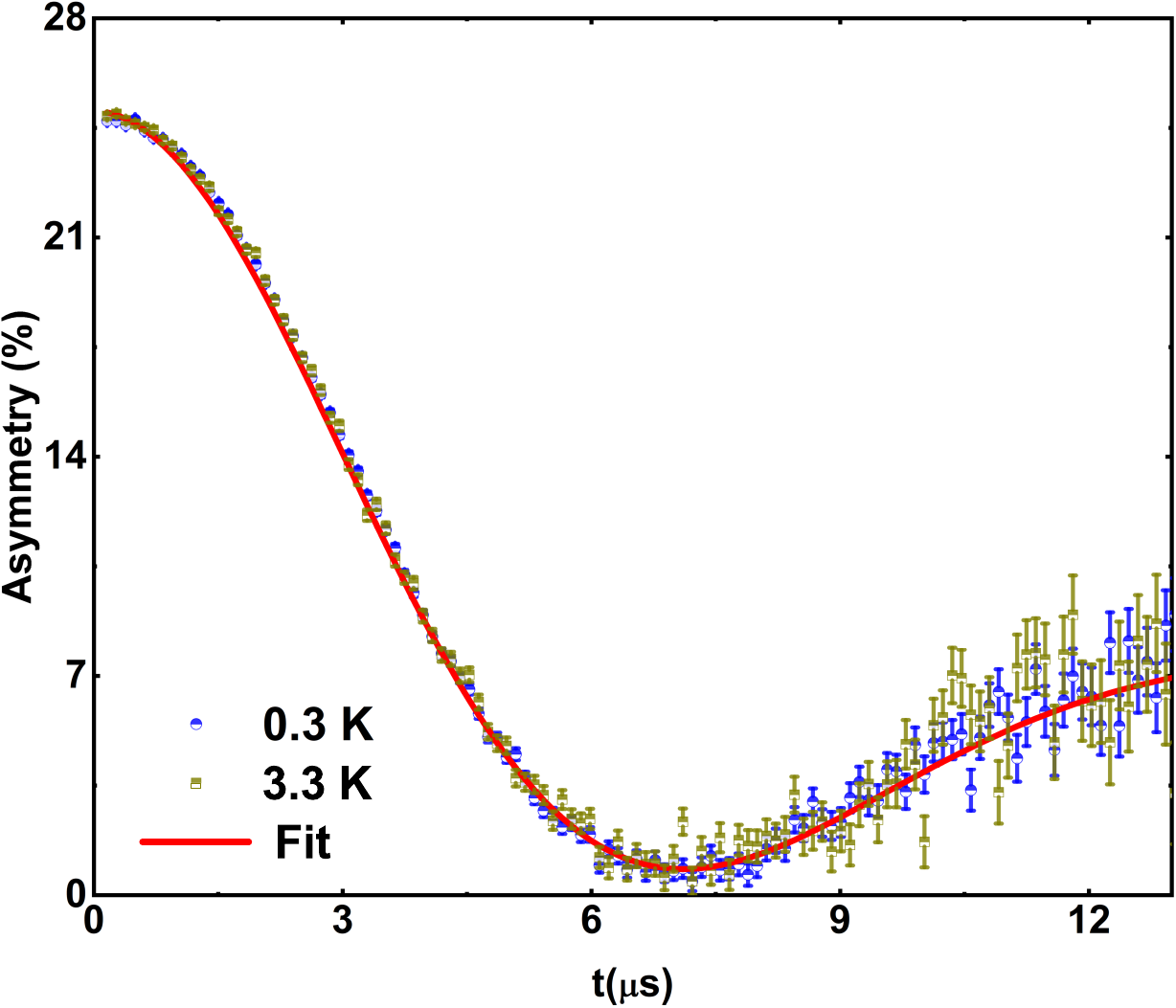}\hfil
\caption{Zero-field $\mu$SR asymmetry spectra for ScRuSi obtained at two distinct temperatures: 0.3\,K (depicted by blue circles) and 3.3\,K (represented by dark yellow squares). The red solid line corresponds to the least squares fit applied to the experimental data. This fit provides a quantitative description of the muon spin relaxation without an external magnetic field, offering insights into the material's behavior across different temperature regimes.}
\label{fig4}
\end{figure}

\subsection{Probing the symmetry of the superconducting energy gap}

To gain insight into the microscopic aspects of the superconducting gap structure, transverse-field muon spin rotation (TF-$\mu$SR) measurements were conducted. Figs.~\ref{fig2}(a)-(b) show the TF-$\mu$SR asymmetry spectra recorded within the vortex state both below and above $T_{\text{C}}$ at an applied magnetic field of 30\,mT. The solid red curves in (a) and (b) represent fits to the data points using the functional form described by Eq. \ref{TF}.~{These fits provide a quantitative understanding of the internal magnetic field distribution of the ground state of ScRuSi.} The observed fast damping of the signal in the superconducting state at 0.1\,K is particularly noteworthy. This effect is indicative of an inhomogeneous field distribution sensed by the muon ensemble, attributed to the formation of vortices in the superconducting material. {The presence of vortices and the behavior of the internal magnetic field under the influence of an external magnetic field, distribution offer valuable information about the superconducting properties of ScRuSi.} Figs.~\ref{fig2}(c) and \ref{fig2}(d) present the corresponding field distributions, determined using the maximum entropy method. In the vortex state, the formation of the flux-line lattice (FLL) results in a non-uniform magnetic field distribution between vortices. This effect is confirmed by the significant damping of the TF-$\mu$SR asymmetry spectra , as depicted in Figures~\ref{fig2}(a)-(b), and the corresponding broadening of the Fourier transform, illustrated in Figures~\ref{fig2}(c)-(d). The presence of a sharp, narrow peak around the applied field, along with an additional broad peak indicative of a magnetic field smaller than the transverse field, is evident in Figure~\ref{fig2}(c). {The origin of the narrow peak is due to muons that did not stop in the sample but stopped in the
Ag-sample holder.} Particularly, this additional peak is absent in Figure~\ref{fig2}(d). Such a distribution is characteristic of type-II superconductivity in the mixed state.

The TF-$\mu$SR asymmetry spectra at all temperatures above and below $T_{\text{C}}$ were analyzed using a Gaussian damped sinusoidal function plus a non-decaying oscillation that contributes to the muons stopping in the silver sample holder~\cite{Bhattacharyya2024, BhattacharyyaThCoC2,CeIr3,bhattacharyya2022}.}  

\begin{equation}
A_\mathrm{TF}(t) = A_\mathrm{sc}\cos\left(\omega_\mathrm{1}t+\varphi\right)\exp\left(-\frac{\sigma_\mathrm{T}^{2}t^{2}}{2}\right)+ A_\mathrm{bg}\cos\left(\omega_2t+\varphi\right).
\label{TF}
\end{equation}

In the analysis of the TF-$\mu$SR data, we obtained $A_\mathrm{sc}~= 89.5\,\%$ as the initial asymmetry originating from the sample, while $A_\mathrm{bg} ~= 10.5\,\%$ accounts for the background arising from muons implanted directly into the silver sample holder, which remains unaffected by depolarization. The muon precession frequencies within the sample and the sample holder are denoted as $\omega_{\mathrm{1}}$ and $\omega_{\mathrm{2}}$, respectively, with $\varphi$ representing a shared initial phase. The total muon-spin relaxation rate within the sample, $\sigma_\mathrm{T}$, is composed of two components: one arising from the inhomogeneous field variation across the superconducting vortex lattice, denoted as $\sigma_\mathrm{sc}$, and the other originating from the normal state, referred to as $\sigma_{\mathrm{nm}} = 0.2195\,\mu s^{-1}$. Importantly, $\sigma_{\mathrm{nm}}$ is considered temperature-independent over the entire studied temperature range and is determined from the spectra measured above $T_\text{C}$. To isolate the superconducting contribution to the depolarization rate, denoted as $\sigma_\mathrm{sc}$, we apply a quadratic subtraction of the temperature-independent nuclear magnetic moment contribution $\sigma_\mathrm{nm}$ from the total depolarization rate. This is expressed as $\sigma_\mathrm{SC} = \left(\sigma_\mathrm{T}^{2} - \sigma_\mathrm{nm}^2\right)^{1/2}$. 

{To understand the details of the superconducting gap structure in ScRuSi, we studied how the magnetic penetration depth/superfluid density changes with temperature using the model}~\cite{Prozorov, Anand2023}:
\begin{align}
\frac{\sigma_\mathrm{sc}(T)}{\sigma_\mathrm{sc}(0)} &= \frac{\lambda_{L}^{-2}\left(T,\Delta_\mathrm{0,i}\right)}{\lambda_{L}^{-2}\left(0,\Delta_\mathrm{0,i}\right)}, \nonumber \\
&= 1 + \frac{1}{\pi}\int_{0}^{2\pi}\int_{\Delta\left(T\right)}^{\infty}\left(\frac{\delta f}{\delta E}\right) \frac{EdEd\phi}{\sqrt{E^{2}-\Delta^2(T,\phi})}.
\end{align}

Here, $f= \left[1+\exp\left(E/k_\mathrm{B}T\right)\right]^{-1}$ represents the Fermi distribution function, and $\Delta(T,\phi) = \Delta_\mathrm{0}\delta(T/T_\mathrm{C})\mathrm{g}(\phi)$ characterizes the temperature- and angle-dependent superconducting gap function. The parameter $\Delta_\mathrm{0}$ denotes the superconducting gap value at 0\,K. The temperature evolution of the superconducting gap is described by $\delta(T/T_\mathrm{C}) = \tanh\left[1.82[1.018\left(T_\mathrm{C}/T-1\right)\right]^{0.51}]$. The angular dependence of the superconducting gap function is denoted by $\mathrm{g}(\phi)$, which equals 1 for an $s$-wave gap \cite{Pang2015, Annet1990}. In Figure \ref{fig3}, we explore the superconducting properties of ScRuSi with a 30 mT applied magnetic field.

In Figure \ref{fig3}a, the temperature dependence of the superconducting normalized depolarization rate, $\sigma_\mathrm{sc}(T)$, is depicted. The solid red line in Figure \ref{fig3}(a) represent the fitted curve using the $s$-wave gap model, offering a detailed description of the superconducting state. The inset in Figure \ref{fig3}a shows the temperature dependence of the total muon-spin depolarization rate, $\sigma_\mathrm{T}$. The experimental data is best represented by a single isotropic $s$-wave gap function of 0.281\,meV. This results in a gap to critical temperature ratio of $2\Delta/k_\mathrm{B}T_\mathrm{C} = 2.71$, indicating that ScRuSi exhibits superconductivity within the weak-coupling limit, { consistent with the heat capacity data~\cite{Ruan2016}}. The agreement between the theoretical model and the experimental data emphasizes the appropriateness of the selected isotropic $s$-wave gap function.

The muon-spin depolarization rate related to the superconducting state, denoted as $\sigma_\mathrm{sc}$, is intricately connected to the {magnetic} penetration depth through the relation~\cite{Brandt2003}
\begin{equation}
\sigma_\mathrm{sc}(T) = 0.0431\frac{\gamma_{\mu}\Phi_{0}}{\lambda_\mathrm{L}^{2}(T)}
\end{equation}
  where $\Phi_0 = 2.609 \times 10^{-15}$\,Wb represents the magnetic flux quantum. The $s$-wave fit yields $\lambda_\mathrm{L}(0)$ = 272\,nm. The London model establishes a direct connection between $\lambda_\mathrm{L}(T)$ and $m^{*}/n_\mathrm{s}$, expressed as $\lambda_\mathrm{L}^2=m^{*}c^{2}/4\pi n_\mathrm{s} e^2$. Here, $m^{*} = (1+\lambda_{\mathrm{e-ph}})m_{\mathrm{e}}$ denotes the effective mass, expressed in units of the electron rest mass $m_{\mathrm{e}}$, and $n_{\mathrm{s}}$ represents the carrier density. $\lambda_{\mathrm{e-ph}}$, {electron phonon coupling constant}, is calculated using the McMillan equation~\cite{McMillan}
\begin{equation}  
\lambda_\mathrm{e-ph} = \frac{1.04+\mu^{*}\ln(\Theta_\mathrm{D}/1.45T_\mathrm{C})}{(1-0.62\mu^{*})\ln(\Theta_\mathrm{D}/1.45T_\mathrm{C})-1.04}
\end{equation}
where $\Theta_{D}$ and $T_\mathrm{C}$ from the Debye temperature and critical temperature, respectively. The estimated superconducting carrier density is $n_\mathrm{s} =5.8 \times 10^{26}$ carriers per m$^{3}$, and the calculated effective-mass enhancement is $m^{*} = 1.51 m_\mathrm{e}$. Similar calculations are detailed in Refs.~\cite{Chia,Amato}. 

Referring to Figure \ref{fig3}b, we study the temperature-dependent field shifts, denoted as $\Delta B = (B_\mathrm{sc} - B_\mathrm{app})/B_\mathrm{app}$, presented as a percentage of the applied field $B_\mathrm{app}$. This data illustrates how the applied magnetic field influences the superconducting state, representing the characteristic diamagnetic shift typically observed in type-II superconductors. The upturn at low temperature in the relative local field shift suggests there may be another contribution to the transverse-field muon spin rotation signal. Currently, the specific origin of this additional contribution to muon spin depolarization is not clear and needs further investigations on the single single of ScRuSi. {The reduction of the diamagnetic shift observed below T $\sim$ 0.4 K in Fig. \ref{fig3}(b) may be due to magnetism, as magnetism appears to alter the diamagnetic shift observed by $\mu$SR at low temperatures in the unrelated superconductor CsCa$_{2}$Fe$_{4}$As$_{4}$F$_{2}$~\cite{Kirschner2018}.}  

\subsection{Time-reversal symmetry}

Time-reversal symmetry (TRS) is a fundamental concept that plays a crucial role in both theoretical and experimental aspects of studying superconductivity. To investigate the time-reversal symmetry in the superconducting state of ScRuSi, we conducted zero-field (ZF) $\mu$SR measurements at temperatures both above and below $T_\text{C}$ (specifically at $T$ = 0.3 and 3.3\,K). The observation of time-reversal symmetry breaking in unconventional superconductors is a key aspect of the study of these materials. It provides valuable clues about the underlying physics, the symmetry of the superconducting order parameter, and the potential emergence of exotic phases with unique properties. TRS breaking has been reported in various ternary equiatomic superconductors, such as (Ta,Nb)OsSi~\cite{Ghosh2022} and (Ta,Nb)RuSi~\cite{Shang2022}. Given the high sensitivity of muons to low magnetic fields, $\mu$SR is a powerful tool for exploring such phenomena. In the ZF-$\mu$SR measurements, the obtained spectra are effectively described by a damped Gaussian Kubo-Toyabe (GKT) depolarization function~\cite{Hayano1979}:
\begin{equation}
A_\mathrm{ZF}(t) = A_\mathrm{sc}G_\mathrm{KT}(t)\exp{(-\lambda_{ZF} t)}+A_\mathrm{bg},
\label{GKTfunction}
\end{equation}
where $A_\mathrm{sc}$ is the initial asymmetry of the sample, $G_\mathrm{KT}(t)$ is the {static Gaussian} Kubo-Toyabe function considering the field distribution at the muon site from both nuclear and electronic moments, $\lambda_\mathrm{ZF}$ represents the damping parameter, and $A_\mathrm{bg}$ denotes the background contribution. {When static electronic moments are absent, the decay of muon ensemble polarization is primarily attributed to randomly oriented nuclear magnetic moments. This decay is commonly described by the static Gaussian Kubo-Toyabe function $G_\mathrm{KT}(t)$.}
\begin{equation}
G_\mathrm{KT}(t) = \left[\frac{1}{3} + \frac{2}{3}\left(1 - \sigma_\mathrm{KT}^{2}t^{2}\right)\exp\left(-\frac{\sigma_\mathrm{KT}^2t^2}{2}\right)\right) 
\end{equation}
The parameters involved in the $G_\mathrm{KT}(t)$ function, namely $A_\mathrm{sc}$, $A_\mathrm{bg}$, and $\sigma_\mathrm{KT}$, are temperature-independent. {$\sigma_\mathrm{KT}$ and $\lambda_{ZF}$ represent the muon-spin relaxation rates originating from the presence of nuclear and electronic moments in the sample, respectively.} Here,{$B_\mathrm{\mu}$} is the width of the local field distribution defined as {${B_\mathrm{\mu}} = \sigma_{KT}/\gamma_\mathrm{\mu}$}, where $\gamma_\mathrm{\mu}/2\pi$ stands for the muon gyromagnetic ratio at 135.53\,MHz/T. This G$_{KT}$ function is particularly relevant in the context of ScRuSi, where the complex nuclear environment, characterized by substantial nuclear spins, contributes to the observed muon-spin depolarization behavior in the absence of an external magnetic field. The manifestation of a G$_{KT}$-shaped function in ScRuSi is expected due to the dense system of nuclear moments with significant nuclear spins. Specifically, the nuclear spins for the relevant isotopes are $I = 7/2$ for $^{45}$Sc, $I = 7/2$ for $^{101}$Ru, and $I = 0$ for $^{28}$Si. The diverse nuclear spins contribute to the local field distribution, creating a distinct G$_{KT}$ behavior observed in the zero-field $\mu$SR spectra of ScRuSi. This characteristic response reflects the complex interplay between the muon spins and the surrounding nuclear moments.

Figure~\ref{fig4} presents a comparison of the ZF-$\mu$SR spectra acquired at 0.3\,K and 3.3\,K. The spectra exhibit identical features, with no discernible muon-spin precession, eliminating the possibility of an internal magnetic field characteristic of magnetically ordered compounds. Notably, the fitting parameters, namely $A_\mathrm{sc}$, $A_\mathrm{bg}$, and $\sigma_\mathrm{KT}$, exhibit no temperature dependence. The absence of significant changes in the relaxation rates between 3.3\,K ($\geq T_\mathrm{C}$) and 0.3\,K ($\leq T_\mathrm{C}$) underscores the stability of these parameters across the superconducting transition. The solid lines in Figure~\ref{fig4} depict the fit to the ZF-$\mu$SR asymmetry data using Eq.~\ref{GKTfunction}. The resulting parameters at 0.3\,K are $\sigma_\mathrm{KT} = 0.2437(2)~\mu \mathrm{s}^{-1}$ and $\lambda_\mathrm{ZF} = 0.0007(3)~\mu \mathrm{s}^{-1}$, while at 3.3\,K, the values are $\sigma_\mathrm{KT} = 0.2431(4)~\mu \mathrm{s}^{-1}$ and $\lambda_\mathrm{ZF} = 0.0005(2)~\mu \mathrm{s}^{-1}$. Importantly, both $\sigma_\mathrm{KT}$ and $\lambda_\mathrm{ZF}$ at temperatures below and above $T_\mathrm{C}$ exhibit agreement within the error threshold. This strongly indicates the preservation of time-reversal symmetry in the superconducting state of ScRuSi.

\begin{figure}[t]
\centering
\includegraphics[width=0.9\linewidth]{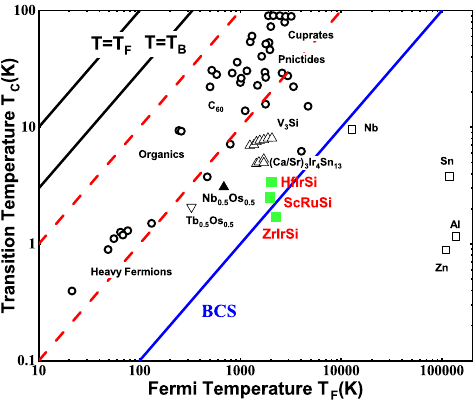}\hfill
\caption{Superconducting critical temperature ($T_\mathrm{C}$) versus the Fermi temperature ($T_\mathrm{F}$) obtained from $\mu$SR measurements in ScRuSi. The shaded area between the dashed red lines delineates the region commonly associated with unconventional superconductors. The position of various conventional BCS superconductors is indicated by the solid blue line for reference. The positions of ScRuSi, HfIrSi, and ZrIrSi place them as conventional superconductors.}
\label{fig5}
\end{figure}

\subsection{Uemura classification scheme}
The Fermi temperature, determined by the expression 
\begin{equation}
T_\mathrm{F} = \frac{\hbar^{2}(3\pi^{2})^{2/3}n_\mathrm{s}^{2/3}}{2k_\mathrm{B}m^{*}}
\end{equation}
This equation involves the carrier density $n_\mathrm{s}$ and the effective mass $m^{*}$ derived from $\mu$SR measurements of the {magnetic} penetration depth. For ScRuSi, the calculated ratio $T_\mathrm{C}/T_\mathrm{F}$ is $2.5/1958 = 0.00128$, categorizing it as a conventional superconductor according to the Uemura classification. The Uemura plot typically reveals that unconventional superconductors display $T_\mathrm{C}/T_\mathrm{F}$ ratios within the range of 0.001-0.1. In contrast, conventional Bardeen–Cooper–Schrieffer (BCS) superconductors occupy the far right of the plot, exhibiting smaller ratios ($1/1000 \geq T_\mathrm{C} / T_\mathrm{F}$). This distinction underscores that unconventional superconductors are characterized by a dilute superfluid, indicating a low density of Cooper pairs, while conventional BCS superconductors manifest a dense superfluid.

In Figure \ref{fig5}, we present a comprehensive plot comparing the critical temperature ($T_\mathrm{C}$) with the Fermi temperature ($T_\mathrm{F}$) derived from $\mu$SR measurements in ScRuSi, framed within the Uemura classification~\cite{Uemura1989, Hillier1997}. This plot not only situates ScRuSi within the broader context of other superconductors but also offers valuable insights into its superconducting behavior. The shaded area between the dashed red lines in the plot signifies the region commonly associated with unconventional superconductors, providing a contextual perspective on ScRuSi's position in the superconducting landscape. The plot includes a solid blue line representing the position of various conventional BCS superconductors, allowing for a clear comparison between conventional and unconventional superconducting behavior. The positions of ScRuSi, HfIrSi, and ZrIrSi are explicitly marked on the plot.

\subsection{First-principles calculations}

\begin{figure}[t]
\centering
\includegraphics[width=\linewidth]{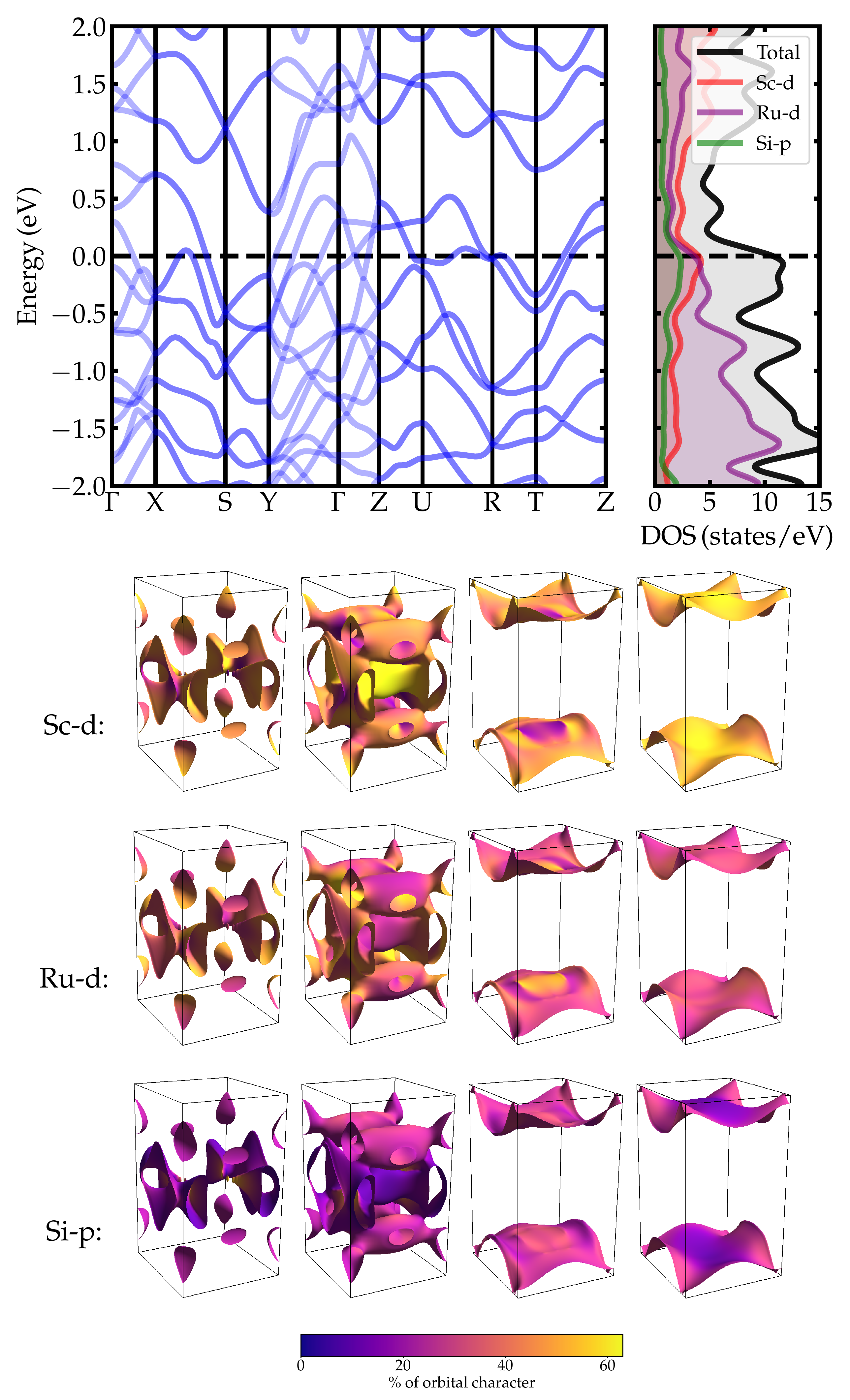}
\caption{Electronic band structure, electronic density of states, and Fermi surface projected onto Sc-$d$, Ru-$d$, and Si-$p$ orbital contributions for ScRuSi.}
\label{fig:electrons}
\end{figure}

The calculated electronic structure of ScRuSi is shown in Figure~\ref{fig:electrons}. ScRuSi is a three-dimensional metal with four bands crossing the Fermi level ($\varepsilon_\text{F}$) and a high carrier density of 10.7 states/eV/unit cell at $\varepsilon_\text{F}$. The Ru-$d$ and Sc-$d$ orbitals are highly hybridized around $\varepsilon_\text{F}$ and represent 70\,\% of the Fermi surface's electronic character. {The expressive amount of Sc-d and Ru-d orbitals across the four different Fermi surface's sheets reveals a strong hybridization and inter-band coupling between the Sc and Ru d shell, as easily verified by inspecting the orbital projection into the Fermi surface in Figure~\ref{fig:electrons}.} Despite its disconnected, complex multiband nature, the ScRuSi Fermi surface presents a homogeneous distribution of the three main orbital characters along the different sheets, which makes it likely to harbor a single, isotropic superconducting gap with conventional pairing symmetry \cite{floris2007, bersier2009,flores2015,heil2017,ferreira2018,bhattacharyya2020,zhao2020,de2021,correa2021,bhattacharyya2022,ferreira2023}, as evidenced in our $\mu$SR measurements. 

\begin{figure}[t]
\centering
\includegraphics[width=\linewidth]{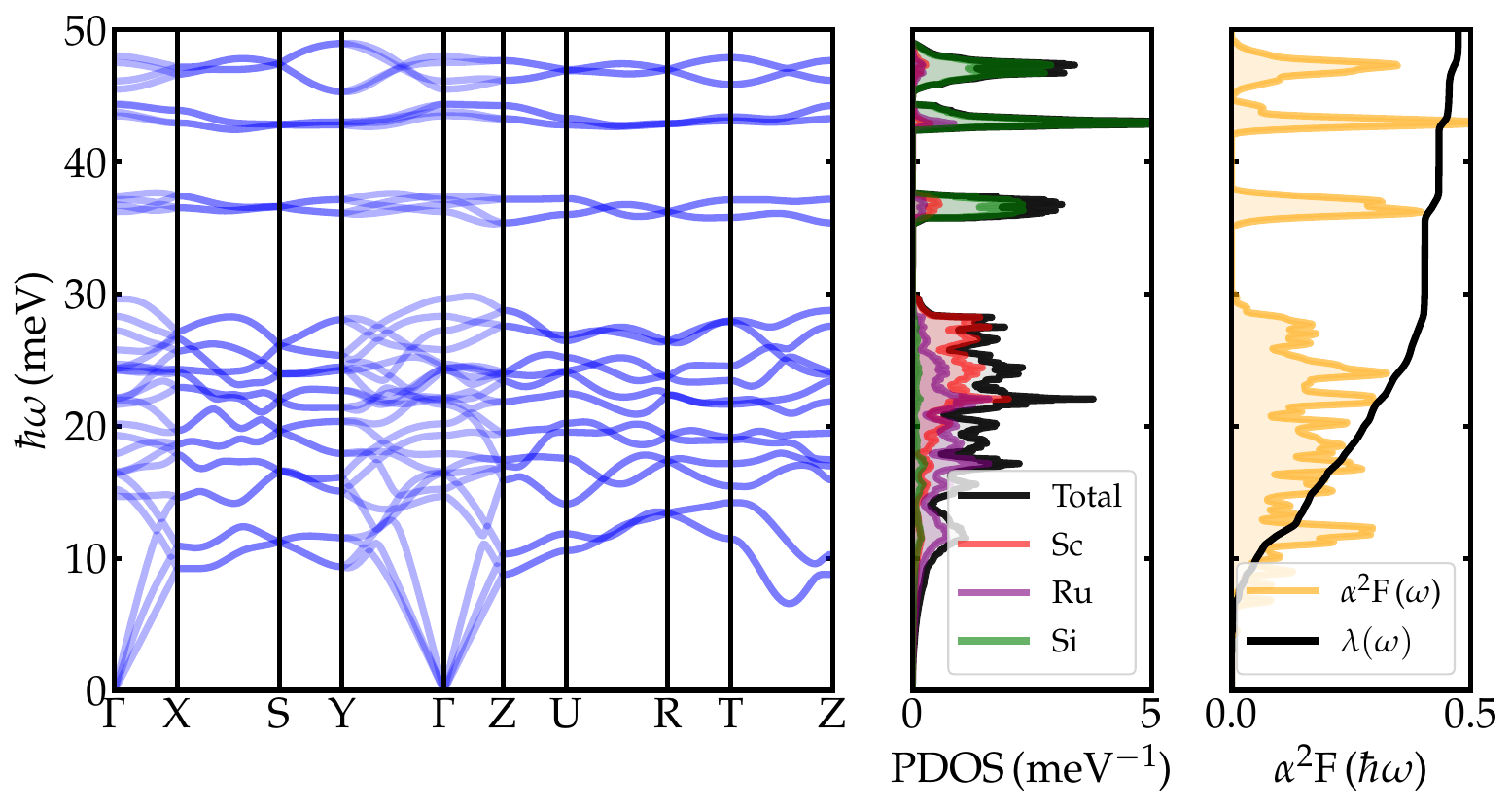}
\caption{Phonons dispersion, phonon density of states, spectral Eliashberg function $\alpha^2F$, and cumulative electron-phonon coupling strength $\lambda_\mathrm{e-ph}$ for ScRuSi.}
\label{fig:phonons}
\end{figure}

Figure~\ref{fig:phonons} shows the phonon bands, the Eliashberg spectral function, $\alpha^2F(\omega)$, and the cumulative electron-phonon coupling parameter, $\lambda_\mathrm{e-ph}(\omega)$, obtained by taking the first reciprocal momentum of the Eliashberg spectral function {calculated
using Quantum Espresso}:
\begin{align}
    \lambda_\mathrm{e-ph} = 2\int d\omega\dfrac{\alpha^2F(\omega)}{\omega}.
\end{align}
Integrating $\alpha^2F(\omega)$ one gets a total $\lambda_\mathrm{e-ph}$ of about 0.48, placing it in the weak coupling regime. This value of $\lambda_\mathrm{e-ph}$ yields an effective-mass enhancement factor of 1.48, in excellent agreement with the value of 1.51 obtained from the muon-spin depolarization rate. A high amount of Si-derived states at the Fermi level would be favorable to superconductivity due to the coupling of high-frequency phonon modes with the electrons near the Fermi surface. Instead, we see a significant contribution to $\lambda_\mathrm{e-ph}$ from Ru low-frequency phonon modes, especially those situated between 5 and 20\,meV in the phonon spectra.

The $\lambda_\mathrm{e-ph}$ parameter of 0.48 gives a predicted $T_{\text{c}}$ of 2.11\,K according to the semi-empirical McMillan-Allen-Dynes equation~\cite{allen1975},
\begin{align}
    T_{\text{c}} &= \dfrac{f_{1}f_{2}\omega_{\text{log}}}{1.20}\exp\left(-\dfrac{1.04(1+\lambda_\mathrm{e-ph})}{\lambda_\mathrm{e-ph} - \mu^{*}(1 + 0.62\lambda_\mathrm{e-ph})}\right),\\
    f_1 &= \left(1 + \left(\dfrac{\lambda_\mathrm{e-ph}}{2.46(1 + 3.8\mu^{*})}\right)^{3/2}\right)^{1/3}, \\
    f_2 &= \left(1 + \dfrac{\lambda_\mathrm{e-ph}^2(\overline{\omega}_{2}/\omega_{\text{log}} - 1)}{\lambda_\mathrm{e-ph}^2 + \left[1.82(1 + 6.3\mu^{*})(\overline{\omega}_{2}/\omega_{\text{log}})\right]^2}\right),
\end{align}
where $f_1$ and $f_2$ are factors depending on $\lambda_\mathrm{e-ph}$, $\mu^{*}$, $\omega_{\log}$, and $\overline{\omega}_{2}$. Here, $\mu^{*}$ is the Morel-Anderson pseudopotential~\cite{morel1962}, which we have set the typical value of 0.1, $\omega_{\log}$ is the logarithmically average phonon frequency, given by
\begin{align}
    \omega_{\log} = \exp\left[\dfrac{2}{\lambda_\mathrm{e-ph}}\int\dfrac{d\omega}{\omega}\alpha^2F(\omega)\log(\omega)\right],
\end{align}
and $\overline{\omega}_{2}$ is the 2nd root of the 2nd moment of the normalized distribution $g(\omega) = 2/\lambda_\mathrm{e-ph}\omega\alpha^2F(\omega)$.
For $\lambda_\mathrm{e-ph}$ close to 0.5, the McMillan-Allen-Dynes formula is quite accurate, therefore, there is no reason for solving the strong-coupling Migdal-Eliashberg equations~\cite{xie2022}. The excellent agreement between the calculated $T_\text{C}$ of 2.11\,K and the obtained experimental $T_\text{C}$ value of 2.5\,K strongly indicates that the single-band electron-phonon $s$-wave pairing is fully compatible for explaining the superconducting properties of ScRuSi. 

\section{Summary}

In this comprehensive investigation, we explored the microscopic superconducting characteristics of $o$-ScRuSi, employing a combination of low-temperature resistivity, heat capacity measurements and muon-spin rotation and relaxation ($\mu$SR) techniques. The findings reveal that ScRuSi is a type-II superconductor, exhibiting a critical temperature ($T_\mathrm{C}$) of $2.5 $\,K.  The temperature dependence of the superfluid density, derived from transverse-field $\mu$SR measurements, is successfully described by an isotropic $s$-wave gap structure, characterized by a ratio of 2$\Delta/k_\mathrm{B}T_\mathrm{C} = 2.71$. The zero field relaxation rate remained consistent above and below $T_\mathrm{C}$, demonstrating the preservation of time-reversal symmetry in the superconducting state of ScRuSi. The theoretical analysis of the electronic and phononic structure supports the conventional electron-phonon pairing mechanism. The predicted $T_\mathrm{C}$ of $2.11$~K within the McMillan-Allen-Dynes theory aligns closely with the experimental observations. By elucidating the microscopic superconducting mechanism in ScRuSi through experiment and theory, this study contributes to the understanding of the interplay between unconventional and conventional superconductivity in equiatomic ternary intermetallics.

\begin{acknowledgments} 

AB expresses gratitude to the Science \& Engineering Research Board for the CRG Research Grant (Grant Numbers: CRG/2020/000698 \& CRG/2022/008528) and CRS Project Proposal at UGC-DAE CSR (Grant Number: CRS/2021-22/03/549). DTA acknowledges the financial support from the EPSRC UK (Grant Reference: EP/W00562X/1), the JSPS for an invitation fellowship, and the Royal Society of London for International Exchange funding. PNF and LTFE would like to thank the São Paulo Research Foundation (FAPESP) for financial support under Grants 2020/08258-0 and 2021/13441-1, and acknowledge the Coordenação de Aperfeiçoamento de Pessoal de Nível Superior – Brasil (CAPES) for financial support under Finance Code 001. PNF and CH acknowledge the Cluster of TU Graz for providing computational resources.
\end{acknowledgments} 

\bibliography{ref}

\end{document}